\documentclass[twocolumn]{aastex63}

\newcommand{\beq}{\begin{equation}}
\newcommand{\eeq}{\end{equation}}
\newcommand{\bea}{\begin{eqnarray}}
\newcommand{\eea}{\end{eqnarray}}

\def\EE{{\cal E}}
\def\BB{{\cal B}}

\received{February 25, 2020}
\revised{April 15, 2020}
\accepted{April 16, 2020}
\shorttitle{Supermassive black holes as possible sources of ultra high energy cosmic rays}
\shortauthors{Tursunov et al.}
\graphicspath{{./}{figures/}}

\begin{document}

\title{Supermassive black holes as possible sources of ultra high energy cosmic rays}

\correspondingauthor{Arman Tursunov}
\email{arman.tursunov@physics.slu.cz}

\author[0000-0001-5845-5487]{Arman Tursunov}
\affiliation{Research Centre for Theoretical Physics and Astrophysics, Institute of Physics, Silesian University in Opava, Bezru{\v c}ovo n{\'a}m.13, CZ-74601 Opava, Czech Republic}

\author[0000-0003-2178-3588]{Zden{\v e}k Stuchl{\'i}k}
\affiliation{Research Centre for Theoretical Physics and Astrophysics, Institute of Physics, Silesian University in Opava, Bezru{\v c}ovo n{\'a}m.13, CZ-74601 Opava, Czech Republic}

\author[0000-0002-4900-5537]{Martin Kolo{\v s}}
\affiliation{Research Centre for Theoretical Physics and Astrophysics, Institute of Physics, Silesian University in Opava, Bezru{\v c}ovo n{\'a}m.13, CZ-74601 Opava, Czech Republic}

\author[0000-0002-4439-9071]{Naresh Dadhich}
\affiliation{IUCAA, Post Bag 4, Ganeshkhind, Pune 411 007, India}

\author[0000-0002-1232-610X]{Bobomurat Ahmedov}
\affiliation{Ulugh Begh Astronomical Institute, Astronomicheskaya 33, Tashkent 100052, Uzbekistan}
\affiliation{Tashkent Institute of Irrigation and Agricultural Mechanization Engineers, Kori Niyoziy 39, Tashkent 100000, Uzbekistan}

%



\begin{abstract}

Production and acceleration mechanisms of ultra-high-energy cosmic rays (UHECRs) of energy $>10^{20}$eV, clearly beyond the GZK-cutoff limit remain unclear that points to exotic nature of the phenomena. Recent observations of extragalactic neutrino may indicate the source of UHECRs being an extragalactic supermassive black hole (SMBH). 
We demonstrate that ultra-efficient energy extraction from rotating SMBH  driven by the magnetic Penrose process (MPP) could indeed foot the bill. We envision ionization of neutral particles, such as neutron beta-decay, skirting close to the black hole horizon that energizes protons to over $10^{20}$eV for SMBH of mass $10^9 M_{\odot}$ and magnetic field of strength $10^4$G. Applied to Galactic center SMBH we have proton energy of order $\approx 10^{15.6}$eV that coincides with the knee of the cosmic ray spectra. We show that large $\gamma_z$ factors of high-energy particles along the escaping directions occur only in the presence of induced charge of the black hole that is known as the Wald charge in the case of uniform magnetic field. It is remarkable that the process neither requires extended acceleration zone, nor fine-tuning of accreting matter parameters. Further, this leads to certain verifiable constraints on SMBH's mass and magnetic field strength as UHECRs sources. This clearly makes ultra-efficient regime of MPP one of the most promising mechanisms for fueling UHECRs powerhouse. 

\end{abstract}

\keywords{black hole physics, ultra high energy cosmic rays, magnetic fields,  magnetic Penrose process, quasars: supermassive black holes}


\section{Introduction} \label{intro}

Recent unprecedented discovery of extragalactic high-energy neutrinos has enabled to pinpoint their source to blazar \citep{2018Sci...361.1378I,2018Sci...361..147.}, which is a supermassive black hole at the distance of $\sim 1.75$Gpc with relativistic jets directed almost exactly towards us. It is generally believed that such neutrinos are tracers of ultra-high-energy cosmic rays (UHECRs). UHECRs are the most energetic among particles detected on Earth, with energy $E > 10^{18}$eV unreachable by current most powerful particle accelerators as LHC with maximum energy $<10^{13}$eV per beam. 
 Constituents of UHECRs were thought to be proton dominated indicated by cosmic ray fluorescence measurements \citep{2010PhRvL.104p1101A,2015APh....64...49A}, although recent observations are suggesting heavier constituents \citep{Aab:2016zth}. For Galactic cosmic rays one should observe anisotropy in arrival direction dominantly on the Galactic plane. As observed by both Pierre Auger Observatory \citep{PAO:2017:sci:,2018ApJ...853L..29A} in the southern hemisphere and Telescope Array in the northern hemisphere \citep{2017APh....86...21A}, UHECRs with energy $>10^{18}$eV are extragalactic with very high confidence level. Spectrum of cosmic rays demonstrate the presence of so-called knees and ankle. The cosmic rays with energy up to $\sim 10^{15.5}$eV (knee) are generally believed to be produced in Galactic supernova explosions, while significant lowering of flux between knee and $10^{18.5}$eV (ankle) suggests change of source of such particles. 
 
The flux of cosmic rays with energies $>5\times 10^{19}$eV is extremely low,  which causes the main difficulty in unveiling their source and its physics. In order to explain the highest-energy cosmic rays several exotic scenarios have been proposed including extra dimensions, violation of Lorentz invariance \citep{Bhattacharjee:1998qc,Rubtsov:2016bea}, existence of new exotic particles \citep{Domokos:1998ry} etc. Among the astrophysical acceleration mechanisms for UHECRs, the relativistic shocks in a plasma of relativistic jets have been previously considered among the most plausible \citep{2000PhST...85..191B}. However, the recent results and estimates may indicate \citep{2018MNRAS.473.2364B} that shock acceleration is not able to account for UHECRs energies above $10^{20}$eV. Therefore, the production and acceleration mechanisms of UHECRs remain unclear. 

Remarkably, a supermassive black hole (SMBH) is the largest energy reservoir in the Universe. By irreducible mass  for a black hole \citep{1971PhRvD...4.3552C} it turns out that a rotating black hole has maximum of 29\%, or $0.29$ of its mass in rotational energy which is available for extraction and can be transformed into energy of accelerated particles. For SMBH with typical mass of $M = 10^9 M_{\odot}$ and dimensionless spin parameter, $a=0.5$, available energy for extraction is of the order of $E_{\rm BH}\approx 10^{74}$eV. It is therefore most pertinent to tap this enormous source most effectively and ultra efficiently. 

In this paper we invoke novel and ultra-efficient regime of magnetic Penrose Process that electromagnetically extracts black hole's rotational energy for accelerating cosmic ray particles to ultra-high energy beyond $10^{20}$eV.

\begin{figure*}
  \centering
  \includegraphics[width=0.6\textwidth]{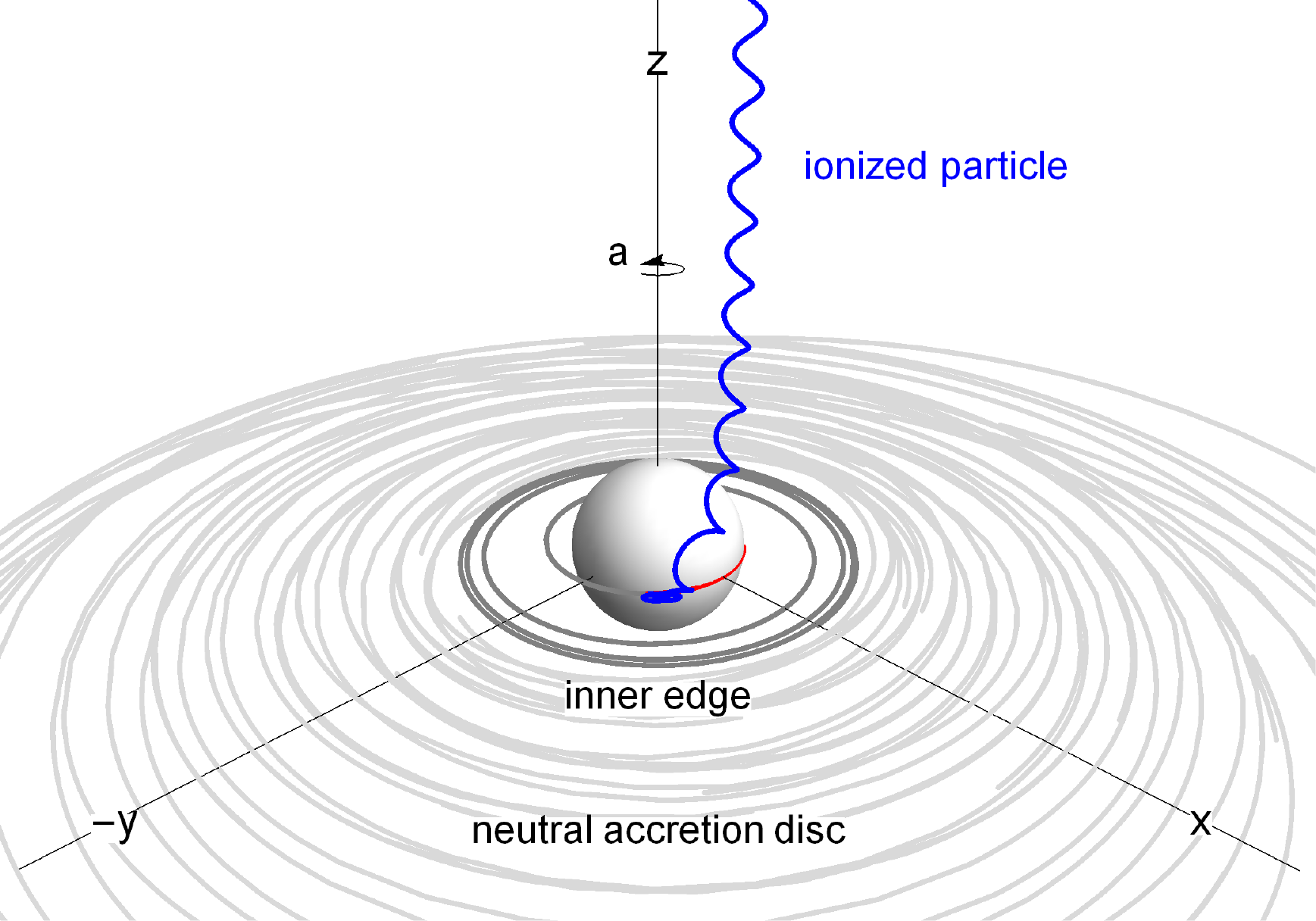}
  \caption{ 
Numerical modelling of the ionization of initially neutral particle (thick grey curve) falling from the inner edge of Keplerian accretion disk onto rotating black hole and resulting escape of positively charged particle (blue curve). Negatively charged fragment after ionization (red curve) collapses into black hole.  The escaping particle after ionization is more likely a positively charged particle due to the presence of more likely positive induced charge of black hole produced by twisting of magnetic field lines \citep{Wald:1974:PHYSR4:,2018MNRAS.480.4408Z}. More detailed numerical analysis of the process is given in section~\ref{sec-accel}. 
}
  \label{fig_schema}
\end{figure*}

\subsection{ { Magnetosphere of astrophysical black hole}} \label{sec-1-1}

In realistic astrophysical scenarios arbitrary electromagnetic field that is present around black hole is weak in a sense that its stress-energy tensor does not contribute to the spacetime geometry. This condition for the magnetic field of the strength $B$ and the black hole of the mass $M$ reads as follows \citep{Gal-Pet:1978:SovJETP:}
\beq 
B << B_{\rm G}=\frac{c^4}{G^{3/2} {M}_{\odot}} \left(\frac{{M}_{\odot}}{M}\right)\approx 10^{19}{ \frac{{M}_{\odot}} {M} }\,{\rm Gauss}\, .
\eeq
Same value of $10^{19}$ holds for the electric field strength  measured in statV$/$cm. Obviously, these conditions are satisfied in all known astrophysical scenarios. Therefore, the spacetime curvature around black hole can be fully described by the standard Kerr metric. Hence arbitrary electromagnetic field surrounding black hole can be considered as a test field in axially symmetric Kerr spacetime background. This weakness of electromagnetic field is compensated by large charge to mass ratio $e/m$ for electrons and protons, whose motion will be essentially affected by magnetic fields already of the order of few Gauss \citep{2017EPJC...77..860K,Tur-Stu-Kol:2016:PHYSR4:}.

First vacuum solution of Maxwell equations in curved spacetime, namely the uniform magnetic field in Kerr metric have been found by \cite{Wald:1974:PHYSR4:}. Later, the effect of plasma in the force-free approximation has been considered in the well-known work of \cite{Bla-Zna:1977:MNRAS:}. Due to lack of direct measurements of exact shapes of the field configurations around realistic black hole candidates, several numerical techniques have been employed that have shown strong connections between the shape of the magnetosphere and the characteristics of accretion process \citep[see, e.g.][]{Punsly:BHmag,Meier:TheEngineParadigm}. In the presence of plasma the magnetosphere has more complicated structure as have been shown by full-fledged general relativistic magnetohydrodynamic (GRMHD) simulations  \citep[see, e.g.][]{Tch:2015:ASSL:,Nak-etal:2018:APJ:,Jan-etal:2018:arXiv:,2019ApJS..243...26P}. Despite that complexity, several analytical solutions to the black hole magnetosphere within the force-free approximation have been proposed \citep[see, e.g.][]{2014MNRAS.445.2500G,2019arXiv190807227G}, where plasma is assumed to be in equilibrium with the electromagnetic field being magnetically dominated.

Although, electromagnetic fields generated by plasma, in general, have non-vanishing electric field components, it is usually expected that electric field is effectively screened by the plasma that makes the system magnetically dominated, i.e. $B^2-E^2>0$. This expectation is satisfied in vacuum \cite{Wald:1974:PHYSR4:} solution everywhere outside the horizon, as well as in  the force-free approach \citep{Bla-Zna:1977:MNRAS:,2014MNRAS.445.2500G}. 
However, in plasma-filled version of the Wald solution, $B^2-E^2$ can turn  negative within the ergosphere of the black hole \citep{2004MNRAS.350..427K}. Special interest is also paid in the literature to the boundary case $B^2=E^2$ that occurs in a  force-free plasma around Kerr black hole \citep{2013CQGra..30s5012B,2007GReGr..39..785M}.

Further, we shall rely on the case in which the magnetic field contribution is dominant everywhere outside the horizon and use the Wald solution for our estimates, in which this condition is satisfied. Near horizon, the strength of induced electric field can be comparable with those of the magnetic field, while decreasing as inverse square of the distance from the black hole.    
Electric field in this case is induced due frame dragging effect of twisting of magnetic field lines that can be seen on the following example. Let us assume that magnetic field is determined by at least one non-zero component of the four-vector potential, so that $A^\phi \neq 0$, which corresponds to the axially symmetric configuration. We should also assume that $A^t = 0$ since any sufficient excess of charge in a plasma will be effectively screened. 
One can see that two co-variant components of the electromagnetic potential are non-vanishing, namely $A_t = g_{t\phi} A^\phi$ and $A_\phi = g_{\phi\phi} A^\phi$. Explicit form of these components in case of the Wald solution we give in details in Section~\ref{sec-accel}. Here we note that the Wald solution refers to a test field in otherwise empty space around a rotating black hole. Introduction of plasma in this scenario does not alter spacetime geometry maintaining test field condition. Plasma like the electromagnetic field would also share the same symmetries of axial symmetry and stationarity. Hence the Wald solution could be forward without much hesitation even in presence of plasma. Here we conclude that the Wald solution can be considered as a simple approximation to the realistic black hole magnetosphere and well suits for both the estimation purposes and understanding of the leading order contributions of corresponding equations. This configuration has been effectively applied in the past for the explanation of various high-energy astrophysical phenomena \citep[see, e.g.][]{2017EPJC...77..860K,2018arXiv180807887L,2019ApJ...886...82R,2019Univ....5..110R,2019arXiv191208174T,2020Univ....6...26S}. Extension of Wald solution to moving black holes in binaries has been obtained by \cite{Mor-Rez-Ahm:2014:PHYSR4:}. Force free approach to similar problem of binary black holes has been studied  by \cite{2012ApJ...754...36A,2012ApJ...749L..32M}.

\section{Basic regimes of energy extraction from rotating black hole} \label{sec-2}

 Magnetic Penrose Process (MPP) has been established in mid 1980's by \cite{Wag-Dhu-Dad:1985:APJ:,Par-etal:1986:APJ:,Bha-Dhu-Dad:1985:JAPA:} as the process allowing the extraction of energy with efficiency exceeding $\eta = 1$ \citep[for a review of early results see][]{Wagh-Dadhich:1989:PR:}. In this section we will show that the efficiency of MPP, under certain conditions, can exceed $\eta \sim 10^{12}$. We define efficiency of energy extraction in a standard manner as the ratio between gain and input energies. In particular, it appears that depending on the initial setup, MPP can work in three basic regimes of efficiencies: low, moderate and ultra. The latter case is able to provide ultra-high energy for charged particles escaping from the vicinity of black holes in a straightforward manner, for characteristic values of magnetic field and even relatively moderate black hole spin. Below, we first discuss the original Penrose process and its relation to other competing mechanisms, giving brief historical remarks and derive its novel, ultra-efficient regime at the end of this section.

The lower limit of MPP refers to the process, originally discovered by \cite{Penrose:1969:NCRS:} in absence of external magnetic field. It is envisaged that a freely falling particle splits into two fragments inside the ergosphere, one of which can attain negative energy relative to observer at infinity while the other respecting energy conservation comes out with energy greater than that of the incident particle. 
Accretion of negative energy particle onto black hole amounts to negative energy flux which is equivalent to extraction of energy from black hole, and the only energy available for extraction is rotational. The maximum efficiency in this regime is only $\eta_{\rm PP} = 0.21$ for 
extremally rotating black hole. Moreover, as shown by \cite{Bar-Pre-Teu:1972:APJ:}, Penrose process requires relative velocity between two fragments to be greater than ${1}/{2} c$ and there is no conceivable astrophysical mechanism that can instantaneously accelerate particles to such high velocity. Thus Penrose process (PP) in absence of electromagnetic interactions was a novel and purely geometric process, but was not astrophysically viable as a power engine for high energy source.

PP was transformed into MPP by taking into account presence of magnetic field produced by surrounding plasma dynamics. Now energy required for particle to ride on negative energy orbit could come from electromagnetic interaction removing all constraints on relative velocity, and thus the process gets revived astrophysically \citep{Wag-Dhu-Dad:1985:APJ:}. Further it was also shown by \cite{Par-etal:1986:APJ:} that its efficiency could exceed $\eta>1$ for discrete particle accretion, a prediction which has been verified by fully relativistic MHD flow simulations in \cite{Nar-McC-Tch:book:2014:}. 

In a plasma setup, another process that could extract energy electromagnetically is the well known \cite{Bla-Zna:1977:MNRAS:} mechanism (BZ). It works on the principle that twisting of magnetic field lines due to frame dragging produces quadrapole electric potential difference between pole and equatorial plane, discharge of which drives energy and angular momentum out from the hole. 
It is generally believed to be leading mechanism for powering relativistic jets observed in variety of black hole candidates. It was shown, however that MPP is more general process than BZ by \cite{2018MNRAS.478L..89D}, since the later requires the presence of threshold magnetic field that is of the order of $\sim 10^4$G, while MPP works for entire range of magnetic field. The latter could be thought of as high magnetic field limit of the former. 
General relativistic MHD simulations by \cite{Las-etal:2014:PYSR4:,Nar-McC-Tch:book:2014:,2015MNRAS.449..316N} have shown 
energy extraction efficiency of this process is moderately high ($\eta \leq 10$), but not ultra-high, exceeding only few hundred percent for polarized plasma in magnetic field. This is the moderate regime of MPP.

On the other hand, there exists the third and the most efficient regime of MPP which can accelerate charged particles to velocities a way higher than one can hope to achieve by above described moderate regime including Blandford-Znajek mechanism (which by its setup uses charged matter only). Here a neutral particle is supposed to split into charged fragments in the ergosphere of rotating black hole in the presence of external magnetic field. As in case of BZ, twisting of magnetic field lines due to frame dragging produces electric field that can be associated to electric charge of the black hole. 
In fact, in both vacuum and plasma cases, the black hole acquires net electric charge proportional to black hole spin \citep[see, e.g.][and references therein]{2018MNRAS.480.4408Z,2018arXiv180807887L,2019Obs...139..231Z,2004MNRAS.350..427K,2001bhgh.book.....P,1997PhyU...40..659B}. 

Neutral particle can reach arbitrarily close to horizon without being influenced by the electromagnetic field, hence split into charged fragments could occur very close to horizon and thereby infalling charged fragment in addition to gravitational/geometric negative energy would have very strong Coloumbic contribution tremendously enhancing quantum of energy being extracted (see, schematic representation of the model in Figure \ref{fig_schema}).  This turns the process "ultra" efficient. 
 For idealized plasma or any other environment (containing charged matter only) as obtaining for BZ, the point of split cannot occur very close to horizon and hence cannot have advantage of tremendous gain of Coloumbic contribution by one of the charged fragments. This is the reason why efficiency of MPP in moderate regime or BZ remains in the moderate range of order of few, as shown e.g. by \cite{2015MNRAS.449..316N} and hence could not reach ultra high range.
Although the proposed model is quite general, further as a particular example we consider beta-decay of neutron (which can appear, e.g., due to nucleosynthesis process in hot and dense plasma of accretion disk \citep{2014A&A...568A.105J}) in dynamical environment of SMBH, from which it follows that proton after decay can naturally reach energy $>10^{20}$eV for the characteristic value of magnetic field of order of $10^4$G, and black hole mass of $10^9 M_{\odot}$.  

 Below we provide main equations supporting discussions given above.  It is generally assumed in all studies of test fields around a rotating black hole that it shares symmetries 
of axial symmetry and stationarity. It should be noted that space around the black hole shares black hole rotation in what is known as the 'frame-dragging' phenomenon. It is therefore electromagnetic field as well as 
plasma share these symmetry properties. This implies that the four-vector potential of the electromagnetic field $A_\mu$ has two non-vanishing covariant components $A_t$ and $A_\phi$  \citep[see, discussion in Sec.~\ref{sec-1-1} and][]{Wald:1974:PHYSR4:}.  The presence of magnetic field modifies the canonical four-momentum  of charged test particles according to $P_\mu = m u_\mu + q A_\mu$, where $m$, $q$ and $u^\mu$ are mass, charge and four-velocity of test particle. Conserved components of the four-momentum, namely energy $E$ and angular momentum $L$, can be written in the form 
\bea
- E &=&  P_t = m u_t + q A_t, \label{KillingEnergy} \\
 L &=& P_\phi = m u_\phi + q A_\phi. \label{KillingAngMom}
\eea
Dynamics of charged particles around Kerr black hole in presence of magnetic field has been widely studied in literature \citep[see, e.g.][and references therein]{2016EPJC...76...32S,Tur-Stu-Kol:2016:PHYSR4:,2018ApJ...861....2T}.

In addition to conservation of energy and angular momentum, the normalization of the four-velocity $u^\alpha u_\alpha = - \delta$ holds for both neutral and charged particles, where $\delta=1$ for massive particle and $\delta=0$ for massless particle. 
Energy of a particle is minimal at the equatorial plane, for which the four-velocity can be rewritten as $u^\alpha = u^t ( 1, v, 0, \Omega)$, where we denote by $v=dr/dt$ and $\Omega = d\phi/dt \equiv u^\phi/u^t$, the radial velocity and the angular velocity of the test particle measured by asymptotic observer, respectively. Substituting this to the normalization condition, we get the expression for $\Omega$  in the form
\bea \label{eq-Omega-gen}
\Omega &=& \frac{1}{D} \left( - \,H g_{t\phi}
\pm \sqrt{u_t^2 \left(H g^2-D g_{rr} v^2\right)} \right),\\
D &=& \delta \, g_{t\phi}^2 + u_t^2 \, g_{\phi\phi}, \quad H = \delta \, g_{tt} + u_t^2, \\
g^2 &=&  g_{t\phi}^2-g_{\phi\phi} g_{tt}, \quad u_t = - \left(  E + q A_t \right)/m.
\eea
The sign in (\ref{eq-Omega-gen}) depends on whether the particle is co-rotating or counter-rotation with respect to locally non-rotating reference frame.  
Possible values of $\Omega$ are restricted by the limit %
 of $u^\alpha$ tending %
 to a null %
vector \citep{Par-etal:1986:APJ:}, i.e. $\Omega_{-} \leq \Omega \leq \Omega_{+}$, where $\Omega_{\pm}$ takes the form 
\beq \label{eq-Omega-pm}
\Omega_{\pm}=\frac{1}{g_{\phi\phi}} \left(-g_{t\phi} \pm \sqrt{g_{t\phi}^2 - g_{tt} g_{\phi\phi}} \right).
\eeq 

Let us now consider split of a particle (1), not necessarily neutral, into two charged fragments (2) and (3) in the black hole ergosphere at the equatorial plane. The conservation laws before and after split can be written in the form
\begin{eqnarray} \label{conser-law}
E_1 = E_2 + E_3, &\,& L_1 = L_2 + L_3, \\
q_1 = q_2 + q_3, &\,& m_1 \geq m_2 + m_3,\\
m_1 \dot{r}_1 = m_2 \dot{r}_2 + m_3 \dot{r}_3, &\,&
0 = m_2 \dot{\theta}_2 + m_3 \dot{\theta}_3, 
\end{eqnarray}
where dots denote derivative with respect to the proper time. 
{
If one of the particles after split, e.g. particle 2, attains negative energy, particle 3 comes out with energy exceeding the energy of incident particle 1 in expense of rotational energy of the black hole. 
Conservation of the four-momentum at the splitting point leads to the following relation  \citep{Bha-Dhu-Dad:1985:JAPA:}
\beq \label{eqmuphi}
m_1 u^{\phi}_1 = m_2 u^{\phi}_2 + m_3 u^{\phi}_3.
\eeq 
Substituting $u^\phi = \Omega ~u^t = - \Omega ~ Y/X$, where $Y = (E+q\,A_t)/m$ and $X = g_{tt} + \Omega ~g_{t\phi}$ one can rewrite (\ref{eqmuphi}) in the form
\beq 
\Omega_1 m_1 Y_1 \frac{X_2 X_3}{X_1} = \Omega_2 m_2 Y_2 X_3 + \Omega_3 m_3 Y_3 X_2.  
\eeq
This equation leads to the final expression for the energy of escaping particle 3, which after several algebraic steps takes the following form
\beq 
E_3=\chi(E_1+q_1A_t)-q_3A_t, 
\eeq \label{E3-mpp}
\beq \label{chi-mpp}
\chi = \frac{\Omega_1 - \Omega_2}{\Omega_3 - \Omega_2}\frac{X_3}{X_1}, \quad X_j = g_{tt} + \Omega_j g_{t\phi}, 
\eeq 
where $\Omega_j = (d \phi / d t)_j$ is an angular velocity of $j$-th particle, given by (\ref{eq-Omega-gen}) with the values limited by (\ref{eq-Omega-pm}). 
}

\begin{figure*}[t]
  \centering
 \includegraphics[width=0.4\textwidth]{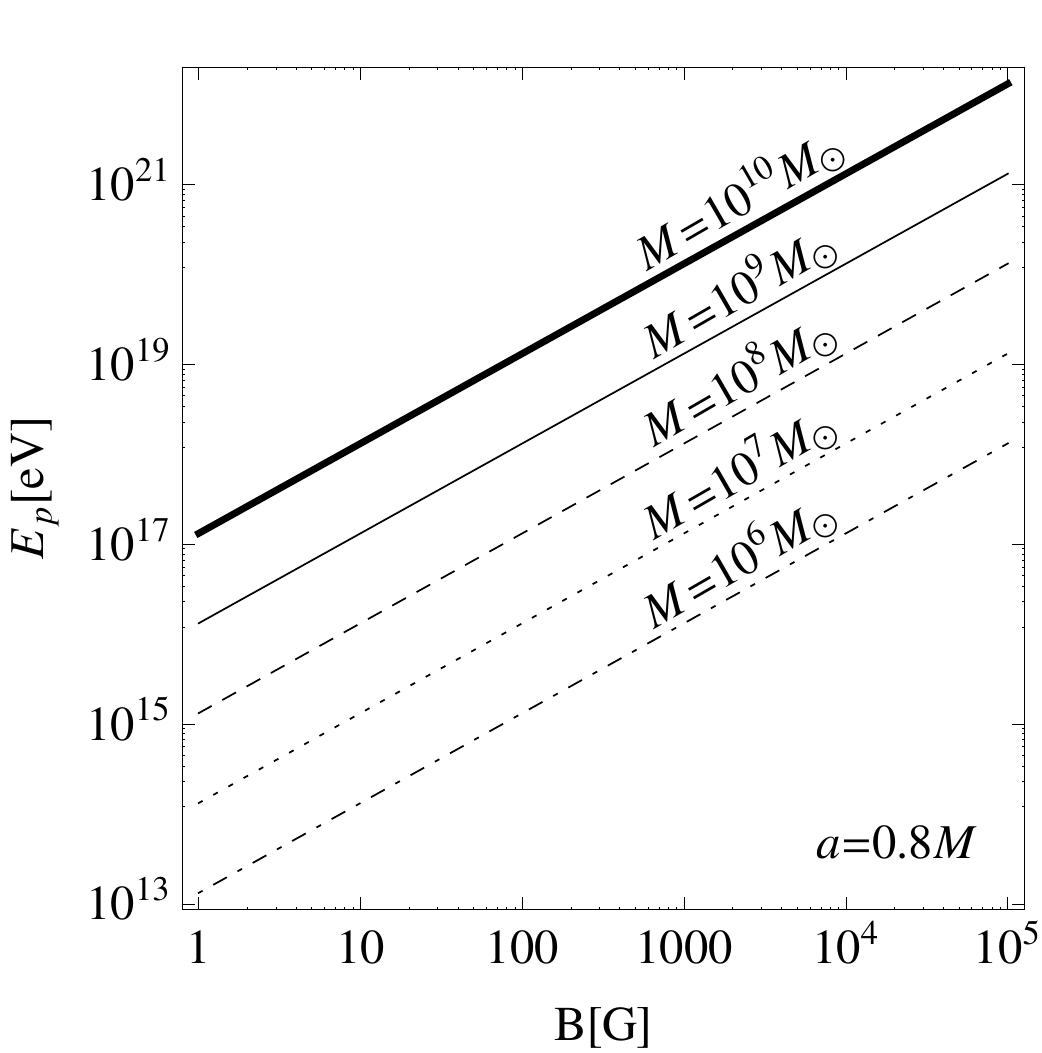}
 \includegraphics[width=0.4\textwidth]{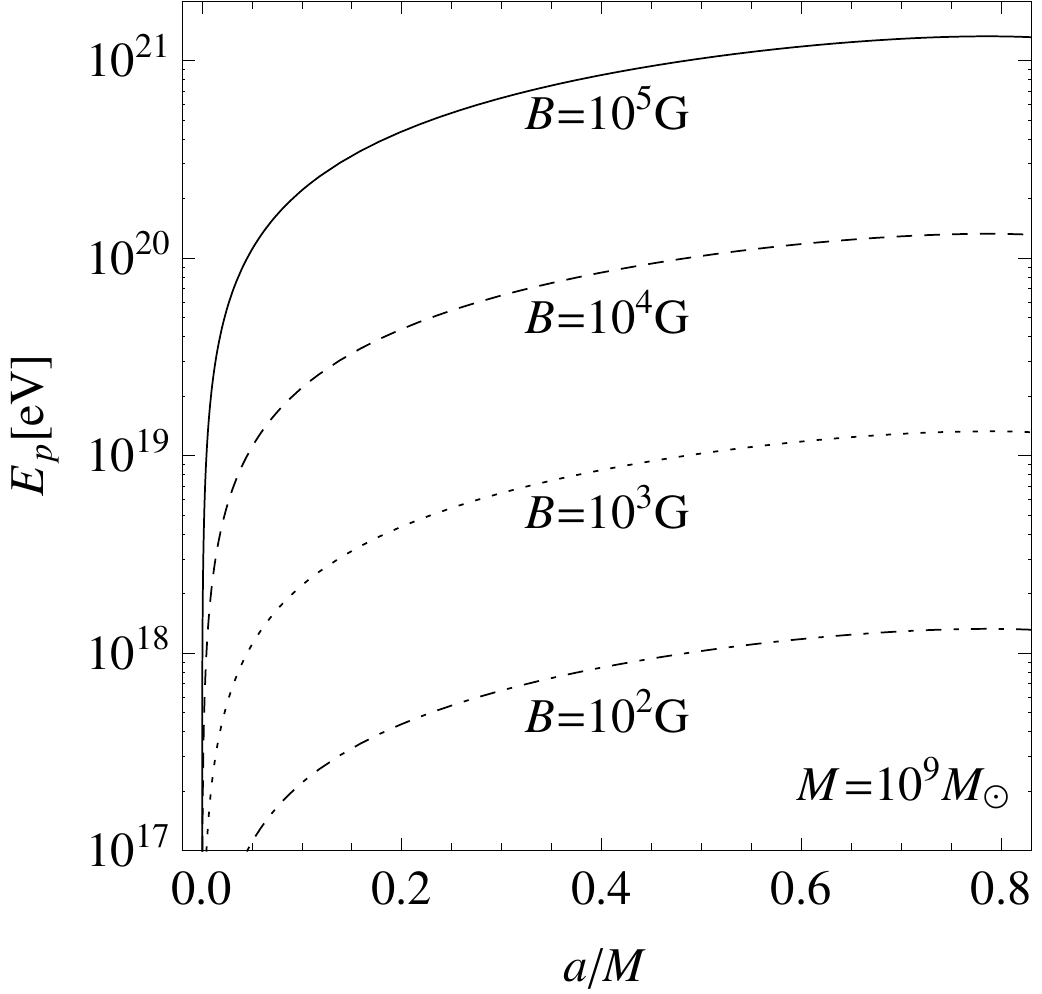}
  \caption{Energy of a proton after neutron beta-decay in dependence on magnetic field (left) and spin (right) for various values of the SMBH's mass and the magnetic field, respectively. }
  \label{fig_3}
\end{figure*}

Let us now define the efficiency of energy extraction as the ratio between gained and infalling energies, i.e. 
\beq \label{eff-general}
\eta = (E_{3} - E_{1})/E_{1} = - E_{2}/E_{1}.
\eeq
If all particles are massive and charged we obtain the following expression for efficiency
\begin{equation} 
\eta = \eta_{\rm PP} + \frac{q_3 A_t - q_1 A_t (\eta_{\rm PP}+1)}{m_1 u_{t1} + q_1 A_t}, \label{efficMPP}
\end{equation} 
where all quantities are calculated at the point of split
and $\eta_{\rm PP}$ is the efficiency of the original \cite{Penrose:1969:NCRS:} process given by purely geometric factors. For the split close to the event horizon, $\eta_{\rm PP}$ takes the form  
 \beq \label{eta-pp}
 \eta_{\rm PP} = \frac{1}{2 a} \, \left( \sqrt{2} \sqrt{ 1 - \sqrt{1 - a^2} } - a \right),   
 \eeq
where $a$ is dimensionless spin of a black hole. This expression coincides with the results of \cite{Penrose:1969:NCRS:} and \cite{Bar-Pre-Teu:1972:APJ:}.

Depending on whether the particles before and after split are charged or neutral one can distinguish three different regimes of efficiencies. In the absence of electromagnetic fields, or if all particles are neutral the expression (\ref{efficMPP}) turns to (\ref{eta-pp}), given by purely geometric factors with the maximum value of $0.21$ for extremally rotating black hole \citep{Penrose:1969:NCRS:}.

If all particles are charged, one can see that in the astrophysically relevant conditions, for elementary particles, such as electrons and protons, the following relation holds: 
 $|\frac{q}{m} A_t| \gg |u_t|$. 
 This inequality implies that in realistic conditions, the motion of charged particle is dominated by the electromagnetic field. When particle is neutral, its specific energy is given by $E/m = -u_t$, which is of the order of unity for a freely falling particle. When the particle is charged, expression for specific energy $E/m = -u_t - \frac{q}{m} A_t$ is sufficiently influenced by the factor $\frac{q}{m} A_t$ that dominates the dynamics due to very large value of the charge to mass ratio $q/m$ for elementary particles. More precise estimate can be obtained numerically \citep[see, e.g.][]{2017EPJC...77..860K}.  
Thus, efficiency of MPP (\ref{efficMPP}) in moderate regime can be reduced to the following simple form 
\beq \label{MPP-mod}
\eta_{\rm mod.} \approx \frac{q_3}{q_1} - 1.
\eeq 
The extraction of energy from black hole occurs when $q_3 > q_1$. Obviously, if particles before and after split are charged, the efficiency remains moderate reaching an order of few in realistic conditions, since a plasma surrounding black hole is usually considered to be neutral.

A situation changes dramatically if the incident particle is neutral ($q_1 = 0$), which splits into two charged fragments. In this case the leading contribution to the efficiency is the third term on the right hand side of Eq. (\ref{efficMPP}). Expression for efficiency in this case takes the form
\beq \label{MPP3}
\eta_{\rm ultra} = \eta_{\rm PP} + \frac{q_3}{m_1} A_t \approx \frac{q_3}{m_1} A_t. 
\eeq 
MPP in this regime is ultra-efficient. Energy of escaping particle, { according to (\ref{eff-general}) is then, given by} 
\beq \label{ener3}
E_3 = \left(\eta_{\rm ultra} + 1 \right) E_1,
\eeq
that can grow ultra-high as we show below.

\section{Maximum energy of a proton from SMBH candidates}

In general, magnetic field has complicated structure in vicinity of the horizon, however in a small fraction of a space where split occurs one can consider the field to be approximately uniform. In this case, known as the Wald solution \citep{Wald:1974:PHYSR4:}, the expression for the ultra-efficiency (\ref{MPP3}) takes the following form 
\beq \label{etaultraeff}
\eta_{\rm ultra} = \frac{1}{2} \left(\sqrt{\frac{r_g}{r_{\rm ion}}} - 1\right) + \frac{q_3 B \, a \, r_g}{2 m_1 c^2} \left(1 - \frac{r_g}{2 r_{\rm ion}}\right),
\eeq
 where $r_{\rm ion}$ is the ionization point (splitting point) of the neutral particle and $r_g = 2 G M/c^2$ is the gravitational radius of a black hole. 
For quantitative estimates, let us consider a neutron beta-decay, 
\beq \label{beta-class}
n^0 \rightarrow p^{+} + W^{-} \rightarrow p^{+} + e^{-} + \bar{\nu}_{\rm e},
\eeq
in vicinity of a SMBH having mass $M$, spin $a$ and magnetic field of strength $B$. 
 Due to large value of the charge to mass ratio for proton, one can see from (\ref{etaultraeff}) that the leading contribution to the efficiency is given by the second term on the right hand side, so that the efficiency can be approximately written as $\eta_{\rm ultra} \approx {e B \, a \, r_g}/{(2 m_{\rm n^0} c^2)}$, where $e$ is the charge of proton and $m_{\rm n^0}$ is neutron mass. From (\ref{ener3}) it follows that the energy of the escaping proton after neutron beta-decay is determined by the relation 
\beq
E_{\rm p^+} = \left(\eta_{\rm ultra} + 1 \right) E_{\rm n^0},
\eeq
were $E_{\rm n^0} \approx 10^9$eV is an energy of a free neutron (mass-energy). 
Substituting here the relation for $\eta_{\rm ultra}$ given by (\ref{etaultraeff}), one can estimate the energy of escaping proton after beta-decay of free neutron as   
\begin{equation} \label{eq1}
\centering
E_{\rm p^+} = 1.7 \times 10^{20} {\rm eV} 
\left( \frac{B}{10^4 {\rm G}}\right) \left(\frac{M}{10^9 M_{\odot}} \right) \left(\frac{a}{0.8} \right),
\end{equation}
predicting energy of proton $E_{\rm p}$ exceeding $10^{20}$ eV for 
$M \sim 10^9 M_{\odot}$ and $B\sim 10^4 G$. Here, we take decay point at $r_{\rm ion} = r_g$, i.e. far enough from the event horizon $r_h = 0.8 r_g$, so that the high-energy particle is allowed to escape to infinity avoiding infinite redshift. In absence of magnetic field, the energy extracting action has to take place very close to horizon for significant efficiency of the process. This is realistically very difficult to sustain and justify. The presence of electromagnetic interaction has completely freed this constraint. 
 In this estimate, we also neglect the effect of antineutrino on the final energy of escaping proton. In Figure~\ref{fig_schema} we depict results of numerical modelling of the ionization of neutral particle skirting in the inner edge of Keplerian accretion disk for schematic purposes. Trajectory of escaping high-energy particle after ionization of freely falling neutral particle from the accretion disk is indicated by blue colour. It is important to note that escaping particle after neutron beta-decay is more likely a proton in the astrophysically favourable cases. This is due to the reason that the Wald charge (or any black hole charge produced by twisting of magnetic field lines) is more likely to be positive in realistic cases \citep[see, e.g. discussions in][]{Wald:1974:PHYSR4:,2018MNRAS.480.4408Z,2019Obs...139..231Z}. More detailed numerical analysis of the process of acceleration is given in the section \ref{sec-accel}.

\begin{figure}
 \includegraphics[width=0.43\textwidth]{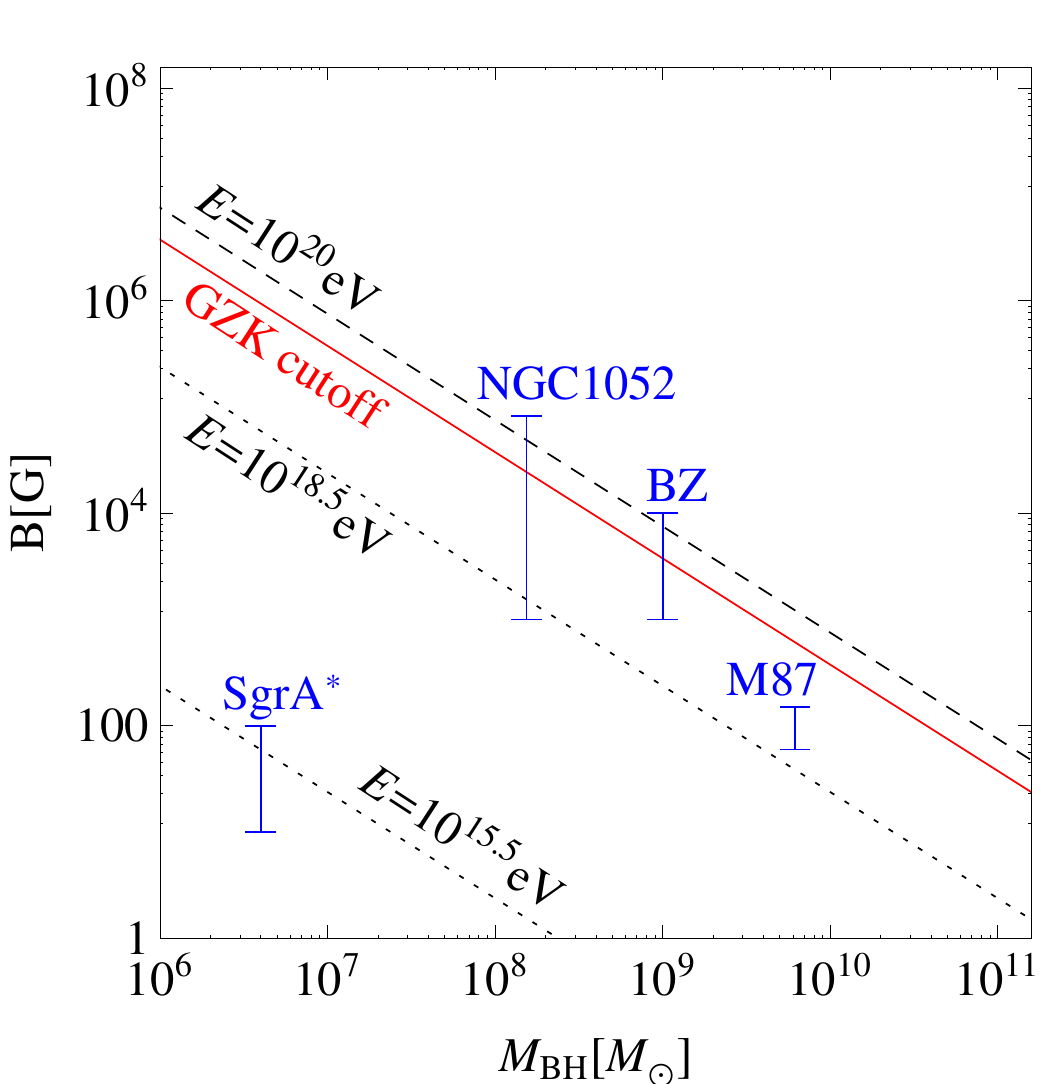}
  \caption{ Constraints on the black hole mass and magnetic field as a source of high-energy protons with various energies: $10^{20}$eV (dashed), GZK-cutoff $10^{19.7}$eV (solid), ankle $10^{18.5}$eV (dotted) and the knee energy $10^{15.5}$eV. Blue vertical lines correspond to SMBH candidates, such as SgrA*, M87 and NGC 1052 with the constraints obtained from observations \citep{2016A&A...593A..47B,2012A&A...537A..52E,2012Sci...338..355D,2015ApJ...803...30K}. The source marked as BZ corresponds to SMBH with mass $10^{9}M_{\odot}$ and magnetic field $10^3 - 10^4$G, consistent with \cite{Bla-Zna:1977:MNRAS:} model of relativistic jets. }
  \label{fig_uhecr}
\end{figure}

The dependence of energy of escaping proton on magnetic field for different black hole masses is given in Figure \ref{fig_3} (left). Proton energy against black hole spin is shown in Figure \ref{fig_3} (right) for SMBH of mass $10^9M_\odot$ and various values of magnetic field. One can see that the process does not require extreme or rapid rotation of the black hole. 

Constraints on magnetic field and SMBH's mass to produce UHECR particles are given in Figure \ref{fig_uhecr}, where several representative SMBH candidates are pointed out. 
Fitting of source candidates given in Figure \ref{fig_uhecr} requires measurements of magnetic fields on the event horizon scales of nearby SMBHs (within approximately 100\,Mpc from the Milky Way, due to GZK-cutoff effect) in addition to mass estimates. Currently, there are only few sources for which magnetic fields are measured on such scales with confidential methods and precisions \citep[we use the estimates obtained in][for corresponding sources]{2016A&A...593A..47B,2012A&A...537A..52E,2012Sci...338..355D,2015ApJ...803...30K}.  
One can also see from Figure \ref{fig_uhecr} that arbitrary SMBH candidate with mass in the range $10^8 - 10^9 M_\odot$ that has been observed with the relativistic jets can serve as a source of protons with energies over $10^{20}$eV, if we assume that the jets are produced in Blandford-Znajek mechanism requiring $10^3 - 10^4$G field (see, bar, indicated as BZ).

Proposed model also gives relatively precise estimate on the maximum energy of protons produced by SMBH in the center of our Galaxy, namely SgrA*, which attributes highly-ordered magnetic field \citep[see, e.g.][]{2015llg..book..391M,Eatough-etal:2013:Natur:,2012A&A...537A..52E} which reaches $10$--$100$G on the event-horizon scales. The mass of SgrA* is estimated to be $\approx 4 \times 10^6 M_{\odot}$  \citep{2017ApJ...845...22P,2012A&A...537A..52E}. Thus the maximum energy of a proton produced after ionization near SgrA*  reaches 
\beq
E_{p^+}^{ \, \rm SgrA^*} = 5 \times 10^{15} {\rm eV}  \left( \frac{B}{10^2 {\rm G}}\right) \left(\frac{M}{4 \times 10^6 M_{\odot}} \right) \left(\frac{a}{0.5} \right). 
\eeq
 This energy remarkably coincides with the knee of cosmic ray energy spectra, above which flux of particles suppresses.

\section{Acceleration and propagation of ionized particles} \label{sec-accel}

\begin{figure*}
  \centering
 \includegraphics[width=\textwidth]{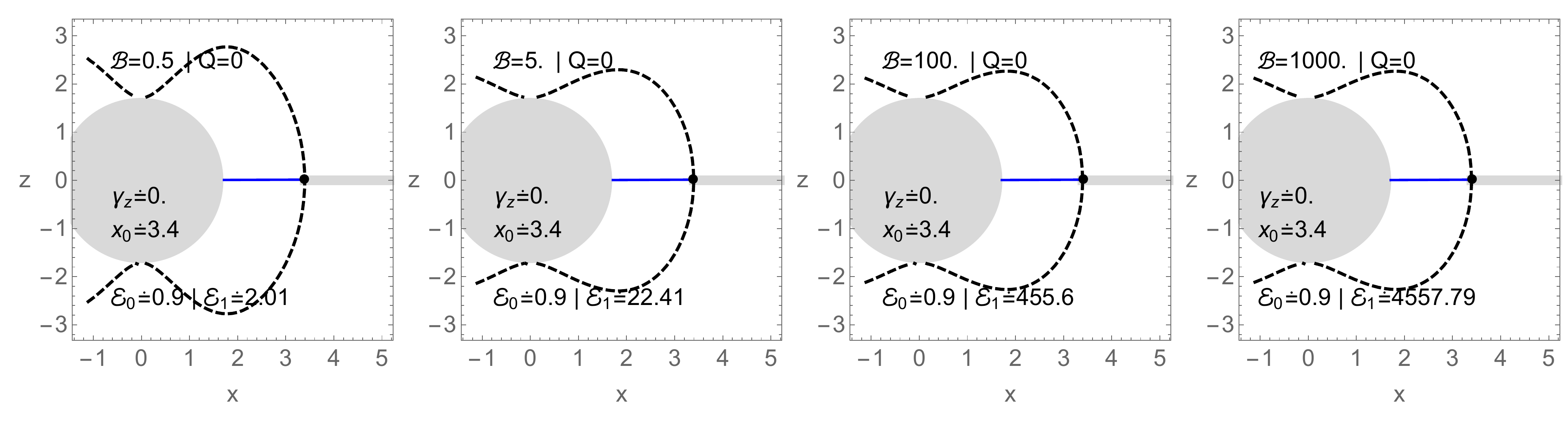}
 \includegraphics[width=\textwidth]{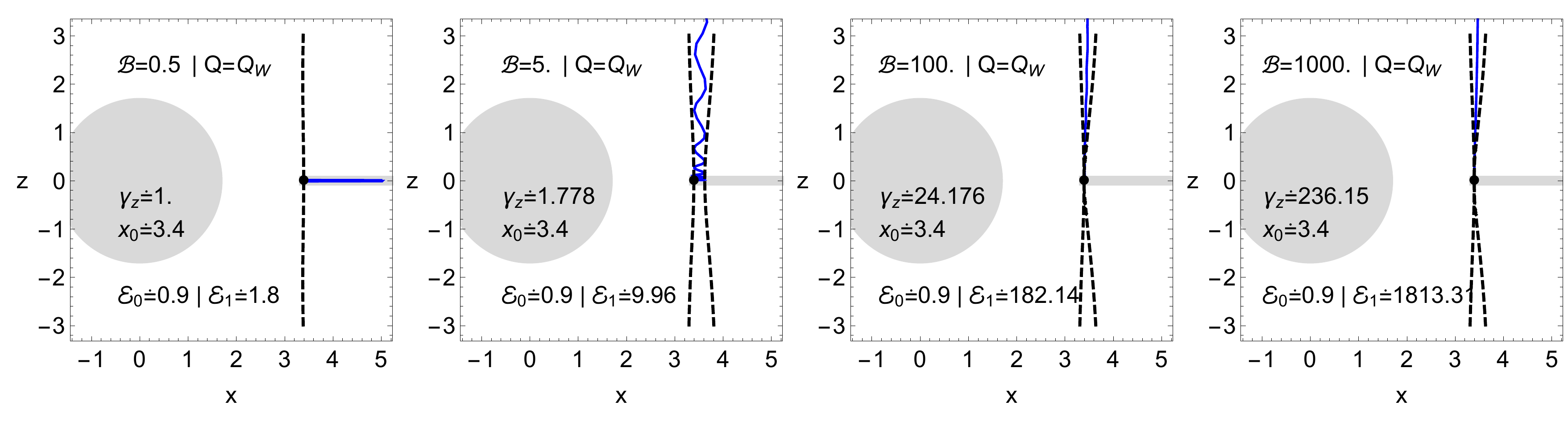}
  \caption{ Comparison of trajectories (blue lines), specific energies ($\EE_1 \equiv E_1/m_1$) and $\gamma_z$-factors of charged particle after ionization of neutral particle with the energy $\EE_0 \equiv E_0/m_0 = 0.9$ in two cases: uncharged black hole (top row) and black hole with induced Wald charge (bottom row). In all plots ionization of neutral particle occurs at the same position $x_0$ (indicated by black dot). Dashed curves represent the boundaries of the motion of charged particle after ionization. Spin of the black hole is chosen to be $a=0.7$. Trajectories are plotted for four values of magnetic parameter $\BB = e G M B/(m_1 c^4)$: $\BB = 0.5$ (first column), $\BB = 5$ (second column), $\BB = 100$ (third column), $\BB = 1000$ (fourth column). Inner edge of the Keplerian accretion disk is shown by grey thick line at the equatorial plane of black hole. In both cases, high-energy particles are produced in the ultra-efficient regime of MPP. Remarkably, the particles escape to infinity only in the presence of induced Wald charge of the black hole (bottom row).}
  \label{fig-accel}
\end{figure*}

\subsection{Escape along the rotation axis}

Boundaries of the motion for charged particles in the axially-symmetric field configurations can be open to infinity along the rotation axis of the black hole, coinciding with the direction of the field lines. Such collimated corridor around axis of rotation of a black hole does not exist for neutral particles. 
 In general, charged particle is allowed to escape to infinity (along the magnetic field lines), if the resulting energy of the particle is greater than its rest energy at infinity. However, in some cases, the Lorentz $\gamma$ factor of charged particle along the rotating axis, i.e. direction of escape remains around unity, although if the energy of the particle is ultra-high.  High-energy charged particle produced at the black hole's equatorial plane can have most of its energy concentrated at the oscillatory (Larmor) energy mode in the equatorial plane, so that the velocity along the axis of rotation (and escape direction) remains zero or moderate. In this section we show that large values of Lorentz $\gamma$ factor of escaping ultra-high-energy particle from the inner regions of the black hole accretion disk may occur only in the presence of the induced charge of the black hole. In the asymptotically uniform magnetic field, this induced black hole charge is known as the Wald charge, arising due to twisting of magnetic field lines by black hole's rotation.

Let us denote the coordinate velocity $v^\mu$, proper velocity $u^\mu$, and Lorentz $\gamma$ factor of the test particle as follows 
\beq
  v^\alpha = \frac{d x^\alpha}{d t}, \quad u^\alpha = \frac{d x^\alpha}{d \tau} = \gamma v^\alpha, \quad \gamma = \frac{d t}{d \tau}. \label{speedDEF1}
\eeq
In the asymptotic limit, i.e. flat spacetime filled by the homogeneous magnetic field one can find that the energy of charged particle measured at infinity is nothing else but 
\beq \label{energyinf}
E_{\infty} = E + q A_t^{\infty},
\eeq 
where $E$ is the integral of motion defined in (\ref{KillingEnergy}) and $A_t^{\infty}$ is the asymptotic value of the time component of the four-vector potential $A_\mu$ of the electromagnetic field that causes the difference between $E_{\infty}$ and $E$. In the case of Kerr black hole immersed into uniform magnetic field of the strength $B$, the asymptotic limit of $A_\mu$ reads as follows 
\beq 
A_{\mu}^{\infty} = \left(- B a, \, 0, \, 0, \, \frac{1}{2} B g_{\phi\phi} \right). 
\eeq
Thus, the Lorentz $\gamma$ factor can be derived in the form 
\beq
 m \gamma = m u^{t} = \frac{d t}{d \tau} = E + q A_{t}^{\infty} = E_{\infty}. \label{gammaDEF}
\eeq
On the other hand, one can see that the energy $E_{\infty}$ can be decomposed into the kinetic energy in escape (vertical) direction, $E_{\rm z}$, and the oscillatory (Larmor) energy, $E_{L}$, in the form \citep[see, for details in][]{2016EPJC...76...32S}
\beq
E_{\infty}^2 = E_{\rm z}^2 + E_{L}^2. 
\eeq
Ejection velocities $u_{\rm z} = u^{\rm z}$ and $v_{\rm z}=v^{\rm z}$ and the corresponding $\gamma_{\rm z}$ factor in the escape direction (coinciding with the rotation axis) can be found in the form 
\beq
m u_{\rm z} = E_{\rm z}, \quad v_{\rm z} = \frac{E_{\rm z}}{E_\infty}, \quad \gamma_{\rm z} = \frac{1}{\sqrt{1-v_{\rm z}^2}} = \frac{E_\infty}{E_L}. \label{velZ}
\eeq
In the flat spacetime, both energies $E_{\rm z}$, $E_{L}$ are conserved, while in the black hole vicinity the transmutation between two energy modes can be observed that can maximize the $\gamma_z$ factor along the black hole rotation axis \citep{2016EPJC...76...32S}. 

Let us now find the condition, in which $\gamma_z$ is maximal. This condition depends on the exact shape of the four-vector potential $A_\mu$, which can have different forms depending on the stage of the black hole accretion. The solution of Maxwell equations for uniform magnetic field of the strength $B$ in the background Kerr black hole spacetime reads \citep{Wald:1974:PHYSR4:} 
\beq 
A_t = \frac{B}{2} \left(g_{t\phi} + 2 a g_{tt}\right), \quad A_{\phi} =  \frac{B}{2} \left(g_{\phi\phi} + 2 a g_{t\phi}\right).
\label{VecPotMax}
\eeq
Here, the rotation of the black hole induces the electric field due to frame-dragging effect that gives a rise to the potential difference between the event horizon and infinity. We can see that the contravariant time components of $A^\mu$ is non-zero, being $A^t = a B$.  Therefore, the solution (\ref{VecPotMax}) for $A_\mu$ in this form causes a selective accretion of charged particles of the same sign into the black hole. Selective accretion into black hole occurs until $A^t = 0$, when remaining non-vanishing component appears to be $A^{\phi} = B/2$.  Covariant components of $A_\mu$ at the final stage of the selective accretion have the following components 
\beq   
A_t = \frac{B}{2} g_{t\phi}, \quad A_{\phi} =  \frac{B}{2} g_{\phi\phi}.
\label{VecPotShort}
\eeq 
At this stage the black hole accretes the charge equal to $Q_W = 2 a M B$ that is known as the induced Wald charge \citep{Wald:1974:PHYSR4:}.  Timescale of selective accretion process is very short for astrophysical black holes. Moreover, induced charge (in different form) should also arise in any other axially symmetric magnetic field configuration different from uniformity. Therefore, one can conclude that any astrophysical black hole candidate possesses non-zero electric charge that is gravitationally weak, however its effect on the charged particles cannot be neglected. Below, we will show that the induced charge of a black hole plays crucial role in the effect of local acceleration of high-energy charged particles produced in the ionization of neutral matter from the inner regions of the black hole accretion flow. In the absence of the induced charge, high-energy particles can also be created within MPP, however, their $\gamma$ factors along the escape direction (coinciding with the rotation and magnetic field axes) remain close to unity.

\subsection{Numerical analysis} 

\begin{figure*}
  \centering
 \includegraphics[width=\textwidth]{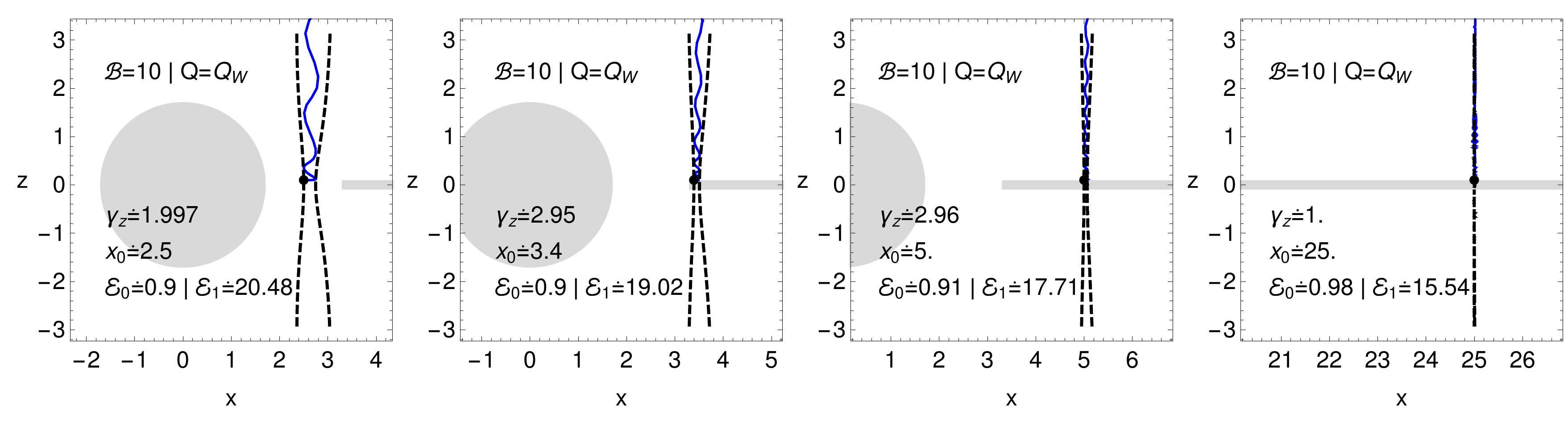}
 \includegraphics[width=\textwidth]{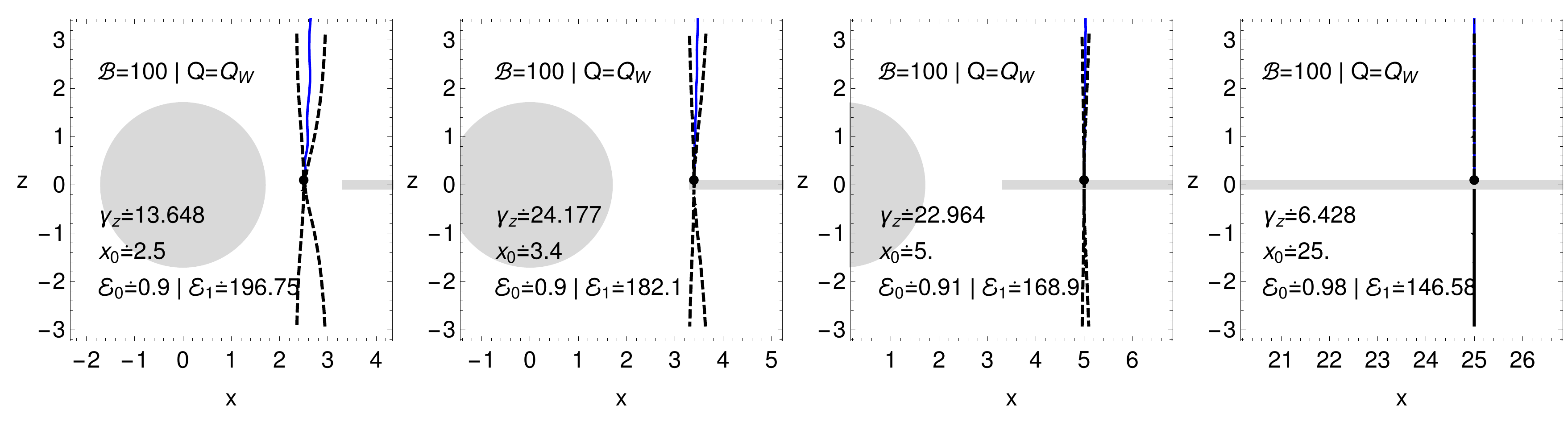}
 \includegraphics[width=\textwidth]{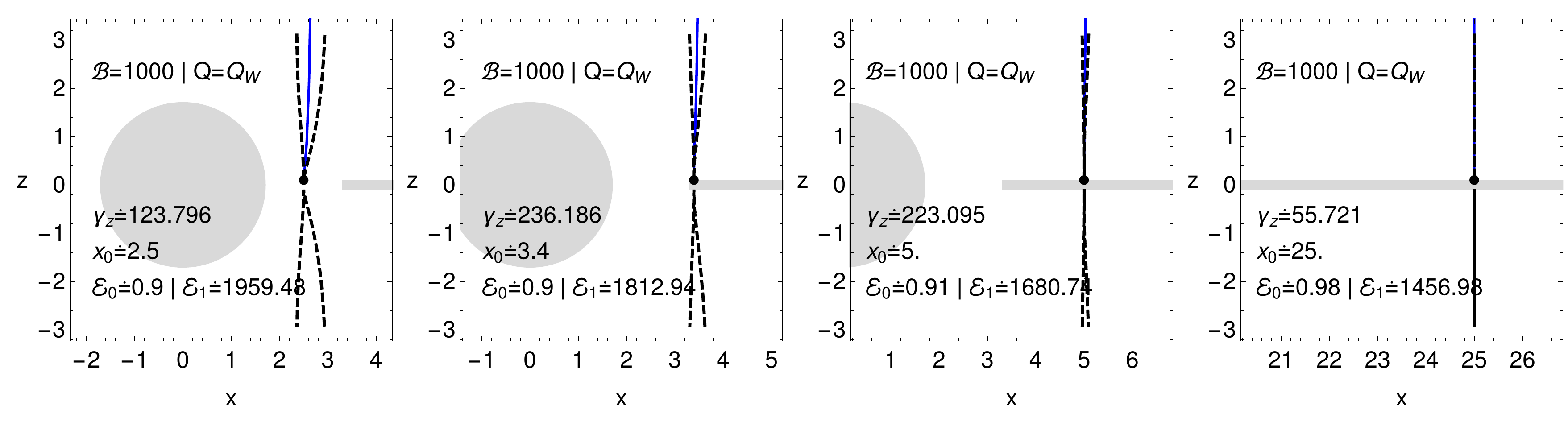}
  \caption{Trajectory (blue lines), specific energy ($\EE_1 \equiv E_1/m_1$), $\gamma_z$-factor of charged particle after ionization of neutral particle with the energy $\EE_0 \equiv E_0/m_0 = 0.9$ at various positions $x_0$ (indicated by black dot) at equatorial plane. Magnetic parameter $\BB = e G M B/(m_1 c^4)$ is chosen as $\BB=10$ (first row), $\BB=100$ (second row), $\BB=1000$ (third row). Dashed curves represent the boundaries of the motion of charged particle after ionization. Spin of the black hole is chosen to be $a=0.7$. Black hole possesses induced charge $Q = Q_W = 2 a M B$. The figure shows that the production of high-energy charged particles does not require the presence of ergosphere at the ionization point.}
  \label{fig-disk}
\end{figure*}

In order to support the above discussions quantitatively, we solve the problem of ionization of initially neutral particles numerically for particular set of initial conditions.  The most general form of the equation of motion of charged particle can be written in the form 
\beq \label{eqmorr}
 \frac{d u^\mu}{d\tau} + \Gamma^{\mu}_{\alpha\beta} u^\alpha u^\beta  = \frac{q}{m} F^{\mu}_{\,\,\,\nu} u^{\nu} + {\cal F}^{\mu}_{\rm rad.}, 
\eeq
where $F_{\mu \nu} = \partial_\mu A_\nu - \partial_\nu A_\mu$ is the tensor of external electromagnetic field and ${\cal F}^{\mu}_{\rm rad.}$ is the radiation reaction force that has sophisticated character, in general \citep[see, e.g.][and references therein]{Poisson:2004:LRR:}. In the astrophysically relevant cases (when considering the motion of charged test particles, such as protons and ions around magnetized Kerr black hole), radiation reaction force can be simplified to the following form \citep[see, details in][]{2018ApJ...861....2T}
\bea \label{rrforce}
{\cal F}^{\alpha}_{\rm rad.} &=&  \frac{2 q^3}{3 m^2} \Big(\frac{D F^{\alpha}_{\,\,\,\beta}}{d x^{\mu}} u^\beta u^\mu \nonumber \\
&+& \frac{q}{m} \left( F^{\alpha}_{\,\,\,\beta} 
F^{\beta}_{\,\,\,\mu} +  F_{\mu\nu} F^{\nu}_{\,\,\,\sigma} u^\sigma u^\alpha \right) u^\mu \Big),
\eea
Radiation losses of cosmic rays due synchrotron radiation in magnetic fields along the propagation distance are discussed in the following subsection. In order to parametrize the equations of motion (\ref{eqmorr}) we introduce the following dimensionless parameter reflecting the relative influence of the Lorentz and gravitational forces
\beq
{\cal B} = \frac{e G M B}{2 m c^4}, 
\eeq
where $M$ is the black hole mass and $e$ and $m$ are the charge and mass of the charged particle. Then, we solve the equations (\ref{eqmorr}) numerically, applying Kerr black hole metric immersed into external uniform magnetic field in the two limiting cases: 
\begin{itemize}
\item uncharged black hole with $A_\mu$ given by (\ref{VecPotMax})
\item black hole with induced Wald charge, $Q=Q_{\rm W} \equiv 2 a M B$, i.e. with $A_\mu$ given by (\ref{VecPotShort}).
\end{itemize}

The results of numerical modelling of MPP, i.e. the ionization of neutral particle near black hole and resulting fate of charged particle is shown in Figure \ref{fig-accel} for uncharged and charged black hole cases. Description of the figure is given in the caption. Here we note that in both cases, ionization of neutral particles lead to the production of high-energy charged particles, however, escape of the particle to infinity (with large escaping velocity) can be observed only in the case of a black hole with induced Wald charge (bottom row of Figure \ref{fig-accel}). Increasing the magnetic parameter $\BB$ leads to increasing the energy, $\gamma_z$ and, importantly, more narrow collimation of escaping charged particles. 
In the absence of the induced charge of the black hole, charged particles perform oscillatory motion around magnetic field lines (Larmor-type of the motion) and this oscillatory energy cannot be effectively transformed into the translational kinetic energy in perpendicular direction. Since the presence of magnetic field always bounds the motion of charged particles in the equatorial plane, final fate of the particle in the absence of induced black hole charge leads to the collapse of the particle into black hole. Therefore, the induced charge of the black hole arising due to twisting of magnetic field lines plays a role of a local accelerator of high-energy charged particles produced in the energy extraction mechanisms, such as MPP. 

It is important to note that the ionization point can be located above the ergosphere of rotating black hole. One can see this in Figure~\ref{fig-disk}, where we plot trajectories of high-energy particles produced at various points of the equatorial plane. One can see that the energy of ionized particle is larger when the ionization point is closer to the black hole. However, the maximal $\gamma_z$-factor (corresponding to velocity in vertical direction) is achieved when the ionization occurs close to the inner edge of accretion disk (ISCO), although the differences between values of specific energy $\EE_1$ and $\gamma_z$-factor of escaping charged particle for ionization points around ISCO and below are not critical. In astrophysical conditions, the charged particles should escape to infinity throughout the disk along the funnels with lower matter density. The highest energy particles should be originated from around ISCO.

In numerical results presented in Figures~\ref{fig-accel} and \ref{fig-disk}, we have used positive values of the magnetic parameter $\BB>0$, which correspond to the motion of positively charged particles, $q>0$, in magnetic field with the field lines oriented in the same direction as the axis of rotation of the black hole. This is the most astrophysically relevant scenario, since in general, magnetic field lines (generated by plasma dynamics co-rotating with a black hole) are supposed to share the axial-symmetry of the black hole at least in its vicinity. This leads to the positive sign of the induced charge of the black hole \citep[see, e.g.][]{Wald:1974:PHYSR4:,2018MNRAS.480.4408Z}.  

In addition to positively charged particle, ionization of neutral particle leads to the creation of negatively charged particle with $\BB<0$, which always has negative energy with respect to observer at infinity within the ergosphere. Outside the ergosphere, the energy of negatively charged particle can be positive, but always lower than the initial energy of incident neutral particle. For large values of magnetic parameter $|\BB| \gg 1$, the energy of negatively charged particle is always negative, due to the energy conservation law given by (\ref{conser-law}). Therefore, negatively charged particle produced in the ionization of initially neutral particle can never escape to infinity remaining bounded or, in most cases, falling into the black hole. This, neutralizes the 'rotationally' induced electric field of the black hole that is equivalent to the extraction of rotational energy of the black hole. In the opposite case of anti-parallel orientation of the magnetic field and rotational axes, induced charge of the black hole is negative. This implies that escaping high-energy particle (produced in the ionization of neutral particle) has to be negatively charged.

In realistic scenarios, the magnetic parameter $\BB$ can be larger for several orders of magnitude, than used in our numerical plots. For protons around typical SMBH of the mass $M = 10^9 M_{\odot}$ and characteristic value of magnetic field $B \sim 10^4 {\rm G}$, parameter $\BB$ has the following value 
\beq
\BB_{\rm SMBH } \approx 2.3 \times 10^{11} \left(\frac{B}{10^4 {\rm G}}\right) \left(\frac{M}{10^9 M_{\odot}} \right).
\eeq
For the best known SMBH candidate, located at the centre of the Milky Way we get the following estimate
\beq \label{est-BBsgra}
\BB_{\rm SgrA^*} \approx 9.4 \times 10^{6} \left( \frac{B}{100 {\rm G}}\right) \left(\frac{M}{4 \times 10^6 M_{\odot}} \right),
\eeq
This implies that the energies, $\gamma_z$-factors and collimation of escaping charged particles in realistic conditions have to be also larger than demonstrated numerically in Figures~\ref{fig-accel} and \ref{fig-disk}. This gives a rise to interpret SMBHs as possible sources of UHECRs, as discussed in the section~\ref{sec-2} with possible candidates given in Figure~\ref{fig_uhecr}, in particular.

 In addition to the induced electric field of the black hole, generated due to frame-dragging effect, electric field can also appear due to charge separation in a plasma surrounding black hole. Circular motion of the plasma of accretion disk around black hole in the presence of the magnetic field (with nonzero component of magnetic field orthogonal to the orbital plane) necessarily leads to the separation of charges in the plasma and resulting non-vanishing component of electric field of plasma. In this case, high-energy charged particles escaping from the inner region of the accretion disk may have additional component of the accelerating force.  Such configuration has been already applied for the investigation of the motion of Galactic centre flare components by \cite{2019arXiv191208174T}. Possible acceleration of cosmic rays by electric field of a plasma surrounding black hole requires further investigation.

\begin{figure}
  \includegraphics[width=0.45\textwidth]{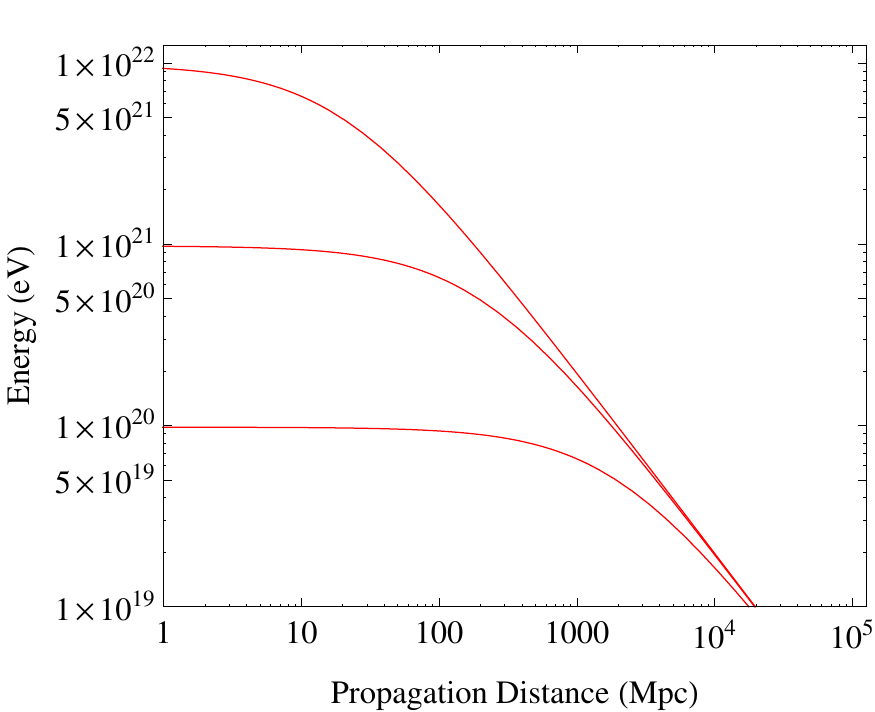}
  \caption{ Mean energy of a proton undergoing synchrotron radiation reaction in a magnetic field of $10^{-5}$G as a function of the propagation distance. }
  \label{fig_synch}
\end{figure}

\subsection{GZK-cutoff and synchrotron losses}

Depending on particle's type and energy, primary UHECR can lose a large part of its energy in the interactions with photons of cosmic microwave background while propagating over distances comparable to size of local cosmological structures. These interactions mainly appear as photo-pion production and force protons with energies above $5\times 10^{19}$eV to lose major part of their energy. Consequently, spectrum of protons shows suppression of flux at these energies, which is known as the GZK cut-off \citep{Greisen:1966jv,Zatsepin:1966jv}. Detection of UHECRs with energies beyond GZK-cutoff imply location of sources within a distance of $\sim 100$Mpc if primary particle is a proton.

On the other hand, inevitable interaction of UHECR with magnetic field along the trajectory can lead to the synchrotron radiation loss. Although suppression of energy of UHECRs in a Galactic and intergalactic magnetic fields is relatively small, UHECRs can loose sufficient amount of their energies in the source regions where magnetic fields can be considerably large. 
For ultra-relativistic particle with charge $q$ and mass $m$ the timescale of synchrotron loss is given by \citep{2018ApJ...861....2T}
\beq \label{tausyn}
\tau \approx \frac{3\, m^3 c^5}{q^4 B^2 f(r)}, \qquad f(r) = 1 - \frac{2 G M}{r c^2}.
\eeq 
Suppression of proton energy on the propagation distance in a magnetic field of $10^{-5}$G is demonstrated in Figure \ref{fig_synch} for different values of initial energy. 
Cubic dependence of the decay timescale (\ref{tausyn}) on the particle mass implies that electrons lose their energy $\sim 10^{10}$ times faster than protons. 
Characteristic timescale of synchrotron energy loss for high-energy electrons propagating in a magnetic field of $10^4$G strength is of the order of $\sim 1$s, against similar timescale for protons that is $\sim 10^{10}$s. Therefore, for typical SMBH with magnetic field of $10^4$G order the primary UHECRs are more plausibly protons or ions, while the decay timescales of electrons are too short for escape from the SMBH vicinity.

\section{Conclusion}

We propose a mechanism which suggests supermassive black holes as sources of ultra-high-energy cosmic rays.
Employing novel, ultra-efficient regime of magnetic Penrose process and ionization of neutral particles particles, such as neutron-beta decay near horizon of spinning black hole we have shown that proton's energy naturally exceeds $10^{20}$eV for SMBH of $10^9M_{\odot}$ and magnetic field of $10^4$G. 
 We list the main advantages of the model as follows:\\
-- clearly predicts SMBHs as the source of highest-energy cosmic rays\\
-- provides verifiable constraints on the mass and magnetic field of the SMBH candidate to produce UHECRs\\
-- operates in viable astrophysical conditions for SMBH with moderate spin and typical magnetic field strength in its vicinity \\
-- does not require extended acceleration zone for particle to reach ultra-high energy, nor the fine-tuning of accreting matter parameters\\ 
-- energy extracting action can take place relatively far from the event horizon without risking the infinite redshift and ultra-efficiency of energy extraction. \\
-- maximum energy of a proton in the process occurring at the Galactic center SMBH  ($10^{15.5}$eV) coincides with the knee of the cosmic ray energy spectra.

The driving engine of the process is in the presence of a  gravitationally induced black hole charge which arises from the magnetic field twist due to black hole rotation in both vacuum and plasma surroundings. Comparing the trajectories of charged particles in the absence and presence of the induced charge numerically, we have shown that production of ultra-high-energy particles after the ionization of neutral particle can be achieved in both cases. However, large velocities in escaping 'vertical' direction can be obtained only in the presence of the induced black hole charge. 

We have shown that the ionization point should not necessarily be within the ergosphere of rotating black hole, although the energy of ionized particle decreases with increasing the distance of ionization point from the black hole. In fact, the maximum escape velocity of charged particle is obtained near the innermost stable circular orbit that is still outside the ergosphere in many cases. 

Our numerical results were obtained in case of vacuum magnetic field, given by the Wald solution. We have also shown that the process should work similarly in any axially-symmetric magnetic field configurations (that shares the symmetries of background Kerr spacetime metric at least near the black hole). Production of UHECRs in plasma MHD case should be tested and we give clear prediction of similar results.

Described mechanism can, in principle, be applied to the neutron stars in which the lower masses are compensated by large values of magnetic fields. Such studies, however, we shall leave for future investigations. 

Since the synchrotron radiation loss of relativistic electrons is of $\sim 10^{10}$ times faster than for protons, heavier constituents of UHECRs seem more plausible in this scenario. 
%

The fit of candidate SMBHs with proposed model requires also the measurements of magnetic fields on the event horizon scale. Nowadays, the number of such precise measurements is very few
and should increase by future global VLBI observations. Constraints on the source candidates with known SMBH masses and magnetic fields are given in Figure~\ref{fig_uhecr}. 

We believe that the proposed model of SMBH as power engine of UHECRs opens up new vista for understanding of this remarkable high energy phenomena as well as of its applications in other similar high energy settings.

\section*{Acknowledgments}

AT acknowledges the International Mobility Project CZ.02.2.69/0.0/0.0/16\_027/0008521 and grateful to the 1st Physics Institute of the University of Cologne and the Inter-University Centre for Astronomy and Astrophysics in Pune for their kind hospitality. 
ZS acknowledges the Albert Einstein Centre for Gravitation and Astrophysics supported by the Czech Science Foundation Grant No. 14-37086G. ND thanks Albert Einstein Institute, Golm for the summer visit. 
BA is supported in part by Projects No. VA-FA-F-2-008 and No. MRB-AN-2019-29 of the Uzbekistan Ministry for Innovative Development, by the Abdus Salam International Centre for Theoretical Physics and by an Erasmus+ exchange grant between SU and NUUz.

\end{document}